# A Theory of Differentiation with Dynamic Clustering

Kunihiko Kaneko and Tetsuya Yomo

[1] Department of Pure and Applied Sciences,
University of Tokyo, Komaba, Meguro-ku, Tokyo 153, JAPAN
[2] Department of Biotechnology, Faculty of Engineering
Osaka University, 2-1 Suita, Osaka 565, JAPAN

**Abstract.** A novel theory for cell differentiation is proposed, based on simulations with interacting artificial cells which have metabolic networks within, and divide into two when the final product is accumulated. Results of simulations with coupled chemical networks and division process lead to the following scenario of the differentiation: Up to some numbers of cells, divisions bring about almost identical cells with synchronized metabolic oscillations. As the number is increased the oscillations lose the synchrony, leading to groups of cells with different phases of oscillations. At later stage this differentiation is fixed in time, and cells spilt into groups with different chemical constituents spontaneously, which are transmitted to daughter cells by cell divisions. Hierarchical differentiation, origin of stem cells, and anomalous differentiation by transplantations are also discussed with relevance to real biological experimental results.

## 1 Introduction

It is often believed that the cell differentiation is completely determined by genes, with some regulatory networks among them[1]. Since genes interact with proteins and other chemicals, however, the differentiation process is not so simple. Indeed, there are some experiments, which cast a question to this widely accepted picture on the cell differentiation: As reported in [2], E. Coli cells with identical genes may split into several groups with different enzymatic activities. Rubin [3], in a series of papers, has shown that the tumor formation strongly depends on the history of the cultivation of cells over several generations, which are not explained by mutations.

In the previous paper[4] we have proposed a novel mechanism which potentially explains the spontaneous cell differentiation based on cellular interactions. The background of this theory lies in recent developments of the clustering theory of globally coupled chaotic elements [5], where chaos leads to the differentiation of identical elements through interaction among them. The relevance of dynamic change of relationships among elements to biological networks has been discussed [6].

In the present paper we extend the previous model[4] to show how cells are differentiated successively into different types. Here we adopt autocatalytic metabolic reaction networks in each cell, while interactions among cells are considered through the medium contacting with cells. We have explicitly included the cell division process, which leads to the increase of the number of cells. Thus the number of equations, consequently the degrees of freedoms of our model increase with time, and our problem provides an example of open chaos discussed earlier [4, 6]. Through our simulations it is shown that the cells lose totipotent ability, as the cells divide, in consistency with well-known fact in the cellular biology. It should be noted that the chemical composition of a cell is inherited by its daughter cells, without imposing any genetic constraints. Furthermore, emergence of stem cells and hierarchical differentiation of cellular types are also discussed.

In our model cells interact through a well stirred medium, and no spatial variation is included. Our results show that differentiation starts by a dynamic, rather than spatial, mechanism in contrast with Turing instability. Indeed our dynamic scenario is consistent with the experimental reports of differentiation in a well stirred medium [2].

## 2 Model

The biochemical mechanisms of the cell growth and division are very complicated, which include a variety of catalytic reactions. The reaction occurs both at the levels of inter- and intra- cells. Here we study a class of models which captures the metabolic reaction and cellular interactions.

Our model for cell society consists of

- Metabolic Reaction Network within each Cell : Intra-cellular Dynamics
- Interaction with Other Cells through Media: Inter-cellular Dynamics
- Cell Division

The basic structure is same as the previous model [4], although the present model includes metabolic network rather than a simple set of reactions, to cope with the complexity in a real cellular system.

**(A) Metabolic Reaction**

First we adopt a set of some chemicals' concentrations as dynamical variables in each cell, and also those in the medium surrounding the cells. We use the following variables; a set of concentrations of chemical substrates $x_i^{(m)}(t)$, the concentration of $m$-th chemical species at the $i$-th cell, at time $t$. The corresponding concentration of the species in the medium is denoted as $X^{(m)}(t)$. We assume that the medium is well stirred, and neglect the spatial variation of the concentration. Furthermore we regard the chemical species $x^{(0)}$ ( or $X^{(0)}$ in the media) as playing the role of the source for other substrates.

The metabolic reactions are usually catalyzed by enzymes, which are inductive and are again synthesized with the aids of other chemicals $x^{(j)}$. Assuming that the dynamics for enzymes is faster, we adiabatically solve the reaction equations of enzyme concentrations, to represent the concentration by those of the substrates ($x^{(j)}$) corresponding to the synthesis [4]. For simplicity we assume that this synthetic reaction is linear in $x^{(j)}$, and adopt the Michaels-Mentens type reaction. Here we use the notation $Con(m,\ell,j)$ which takes the value 1 when there is a metabolic path from the chemical $m$ to $\ell$ catalyzed by the chemical $j$, and takes 0 otherwise. In other words, a metabolic path $x^{(m)} \rightarrow x^{(\ell)}$ produces $x^{(\ell)}$, with the aid of the chemical $j$ when $Con(m,\ell,j)=1$. Here the choice of connected paths depends on each chemical $\ell$, and generally there can be several paths for the production of $\ell$. Thus the reaction from the chemical $m$ to $\ell$ aided by the chemical $j$ leads to the term $e_1 x_i^{(j)}(t) x_i^{(m)}(t)/(1+x_i^{(m)}(t)/x_M)$, where $x_M$ is a parameter for the Michaels-Mentens form. The coefficients for chemical reactions are taken to be identical ($e_1$) for all paths.

In addition, we assume that there is a path to the final product, from all $x^{(k)}$, leading to a linear decay of $x^{(k)}$, with a coefficient $\delta$. Summing up all these processes, we obtain the following contribution of the metabolic network to the growth of $x_i^{(\ell)}$ ( i.e., $dx_i^{(\ell)}(t)/dt$);

$$Met_i^{(\ell)}(t) = e_1 x_i^{(0)}(t) x_i^{(\ell)}(t) + \sum_{m,j} Con(m,\ell,j) e_1 x_i^{(j)}(t) x_i^{(m)}(t)/(1+x_i^{(m)}(t)/x_M)$$

$$-\sum_{m',j'} Con(\ell,m',j') e_1 x_i^{(\ell)}(t) x_i^{(j')}(t)/(1+x_i^{(\ell)}(t)/x_M) - \delta x_i^{(\ell)}(t), \qquad (1)$$

where we note that the two terms with $\sum Con(\cdots)$ represent metabolic paths coming into $\ell$ and out of $\ell$ respectively.

When $m=\ell$, the reaction is regarded as autocatalytic, in the sense that there is a positive feedback to generate the chemical $k$. (In general, it is natural to assume that a set of chemicals work as an autocatalytic set.) Later we will study the case only with autocatalytic reactions, in a more detail.

**(B) Active Transport and Diffusion through Membrane**

A cell takes chemicals from the surrounding medium. Thus cells interact with each other indirectly through the medium. It is expected that the rates of chemicals transported into a cell are proportional to their concentrations outside. Further we assume that this transport rate also depends on the internal state of a cell. Since the transport here requires energy [1], the transport rate depends on the activities of a cell. To be specific, we choose the following form;

$$Transp_i^{(m)}(t) = (\sum_{k=1} x_i^{(k)}(t)) X^{(m)}(t) \qquad (2)$$

The summation ($\sum_{k=1} x_i^{(k)}(t)$) is introduced here to mean that a cell with more chemicals is more active. We choose the above bi-linear form for simplicity, although a nonlinear dependence on $\sum_{k=1} x_i^{(k)}(t)$ with a positive feedback effect leads to qualitatively similar results. Besides the above active transport, the chemicals spread out through the membrane by a normal diffusion process written as

$$Diff_i^{(m)}(t) = D(X^{(m)}(t) - x_i^{(m)}(t)) \quad (3)$$

Combining the processes (A) and (B), the dynamics for $x_i^{(m)}(t)$ is given by

$$dx_i^{(0)}(t)/dt = -e_1 x_i^{(0)}(t) \sum_\ell x_i^\ell(t) + Transp_i^{(0)}(t) + Diff_i^{(0)}(t), \quad (4)$$

$$dx_i^{\ell)}(t)/dt = Met_i^{(\ell)}(t) + Transp_i^{(\ell)}(t) + Diff_i^{(\ell)}(t), \quad (5)$$

Since the present processes are just the transportation of chemicals through membranes, the sum of the chemicals must be conserved. The dynamics of the chemicals in the medium is then obtained by converting the sign, i.e.,

$$dX^{(m)}(t)/dt = -\sum_{i=1}^{N}\{Transp_i^{(m)}(t) + Diff_i^{(m)}(t)\}, \quad (6)$$

where $N$ is the number of cells, which can change in time by cell divisions.

Since the chemicals in the medium can be consumed with the flow to the cells, we need some flow of chemicals (nutrition) into the medium from the outside. Here only the source chemical $X^0$ is supplied by a flow into the medium. By denoting the external concentration of the chemicals by $\overline{X^0}$ and its flow rate per volume of the medium by $f$, the dynamics of source chemicals in the media is written as

$$dX^{(0)}(t)/dt = f(\overline{X^0} - X^0) - \sum_{i=1}^{N}\{Transp_i^{(0)}(t) + Diff_i^{(0)}(t)\}. \quad (7)$$

**(C) Cell Division**

Through chemical processes, cells can replicate. For the division, accumulation of some products is required. In our model the final product, generated from all chemical species, is assumed to act as the chemical for the cell division. (This final product can be regarded as DNA).

$$\int_{t_0(i)}^{T} dt \sum_k \delta \times x_i^{(k)}(t) > R \quad (8)$$

is satisfied, where $R$ is the threshold for the cell replication. Here again, choices of some other division conditions can give qualitatively similar results as those to be discussed. We note that the division condition satisfies an integral form representing the accumulation.

When a cell divides, two almost identical cells are formed. The chemicals $x_i^{(m)}$ are almost equally distributed. "Almost" here means that each cell after a division has $(\frac{1}{2} + \epsilon)x_i^{(m)}$ and $(\frac{1}{2} - \epsilon)x_i^{(m)}$ respectively with a small "noise" $\epsilon$, a random number with small amplitude, say over $[-10^{-3}, 10^{-3}]$. We should note that this inclusion of imbalance is not essential to our differentiation. Indeed any tiny difference is amplified to yield a macroscopic differentiation. It should be noted that for simplicity the volume of a cell is approximated to be constant except for a short span for the division. During the short span for the division, the volume is twice and thus the concentration is made half in the above process.

## 3 Few remarks on internal dynamics

Before presenting the dynamics of cell society, let us briefly describe the nature of metabolic reaction given by (1). Roughly speaking, the dynamics strongly depends on the number of autocatalytic paths. If the number is large, only few chemicals are activated, and all other chemicals' concentrations vanish. In this case no metabolic paths are active, since the ongoing reaction is just the chemical 0 (source)$\rightarrow x^{(k)} \rightarrow$ the final product, without any reactions $x^{(k)} \rightarrow x^{(\ell)}$. On the other hand, when the number of autocatalytic paths is small, many chemicals are generated, but their concentrartions do not oscillate and are fixed in time. When the number of autocatalytic paths is medium, non-trivial metabolic reactions appear. Some, (not necessarily all) chemicals are activated. The concentrations of chemicals oscillate in time, which often shows a switching-like behavior: That is, chemicals switch between low and high values successively. This type of switching behavior is also seen in the randomly connected Lotka-Volterra equations as saddle-connection-type dynamics [7].

In the present paper we discuss cases with a medium number of autocatalytic paths, since they lead to non-trivial metabolic oscillations.

## 4 Proposed Scenario on cell differentiation

We have carried out several simulations of our model with $k = 8$, 16 or 64, with a variety of randomly chosen metabolic networks with connections from 2 to 6 per chemicals. Through these simulations, we propose the following scenario of the cell differentiation. Here we describe our scenario together with numerical results for a given network, although simulations with a variety of metabolic networks support the scenario rather well.

**(1)Metabolic Oscillation of chemicals and Synchronized Division**

Chemical concentrations within each cell oscillate in time by the metabolic reaction process, which provides the basis of the following differentiation process. Up to some number of cells, the oscillations are coherent, and all cells have almost same concentrations of chemicals. Accordingly, the cells divide almost simultaneously, and the number of cells increase as 1,2,4,8. It is interesting to note that cells in most real organisms are not differentiated up to some number of divisions.

**(2) Clustering by Phases of Oscillations**

As the division process proceeds, the metabolic oscillation starts to lose its synchrony. Cells often separate into several groups with distinct phases of oscillations, while the synhorny is preserved within each group of cells. Thus the differentiation sets in. At this stage, however, the differentiation is not yet fixed. In other words, only the phases of oscillation are different by cells, but the temporal averages of chemicals, measured over some periods of oscillations, are almost identical per cells. As has been discussed[4], this temporal clustering corresponds to time sharing for resources, since the ability to get them depends on the chemical activities of cells. In Fig.1a), the temporal averages and snapshot values of some chemicals are plotted in the order of the birth time of cells. Here the averages are almost identical, but the snapshot chemical values starts to show slightly different values. The difference here, indeed, is a trigger to the fixed differentiation at the next stage.

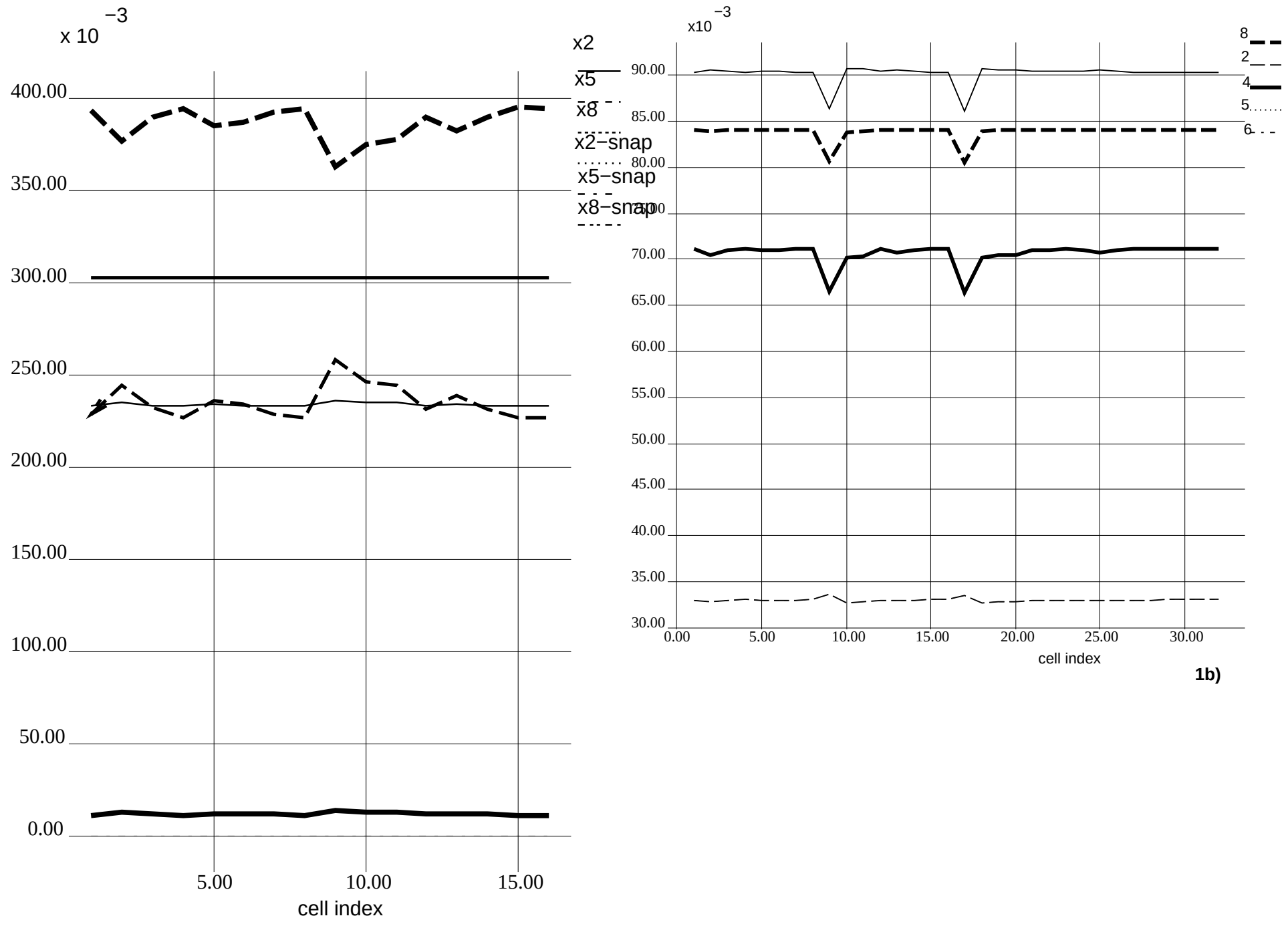

1a)

**(3) Fixed Differentiation**

After some divisions of cells ( for example, at the stage of 32 cells) differences in chemicals start to be fixed by cells. When we measure the average chemicals over periods of oscillations, amounts of chemicals as well as the ratios of chemicals differ by cells. Thus cells with different chemical compositions are generated. This differentiation of cells is not only for the strength of activities [4], but also for the compositions of chemicals. In Fig.1b)-e), we have plotted the averages of chemicals for different temporal regimes. Distinct two groups of cells are created when the cell number is 32 in the figure. ( see the item (6) for the further differentiation at a later stage).

It is noted that the phase difference still remains by each group. Thus there

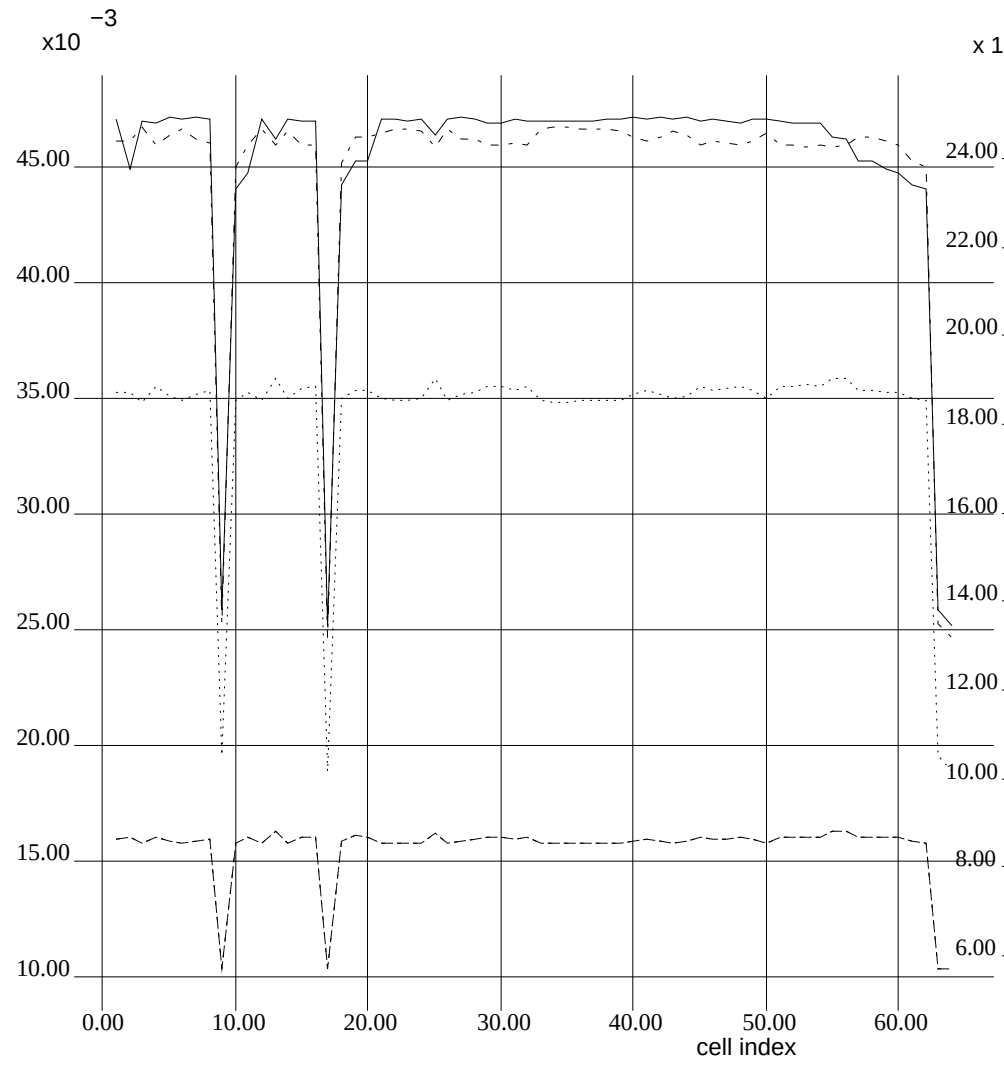

1c)

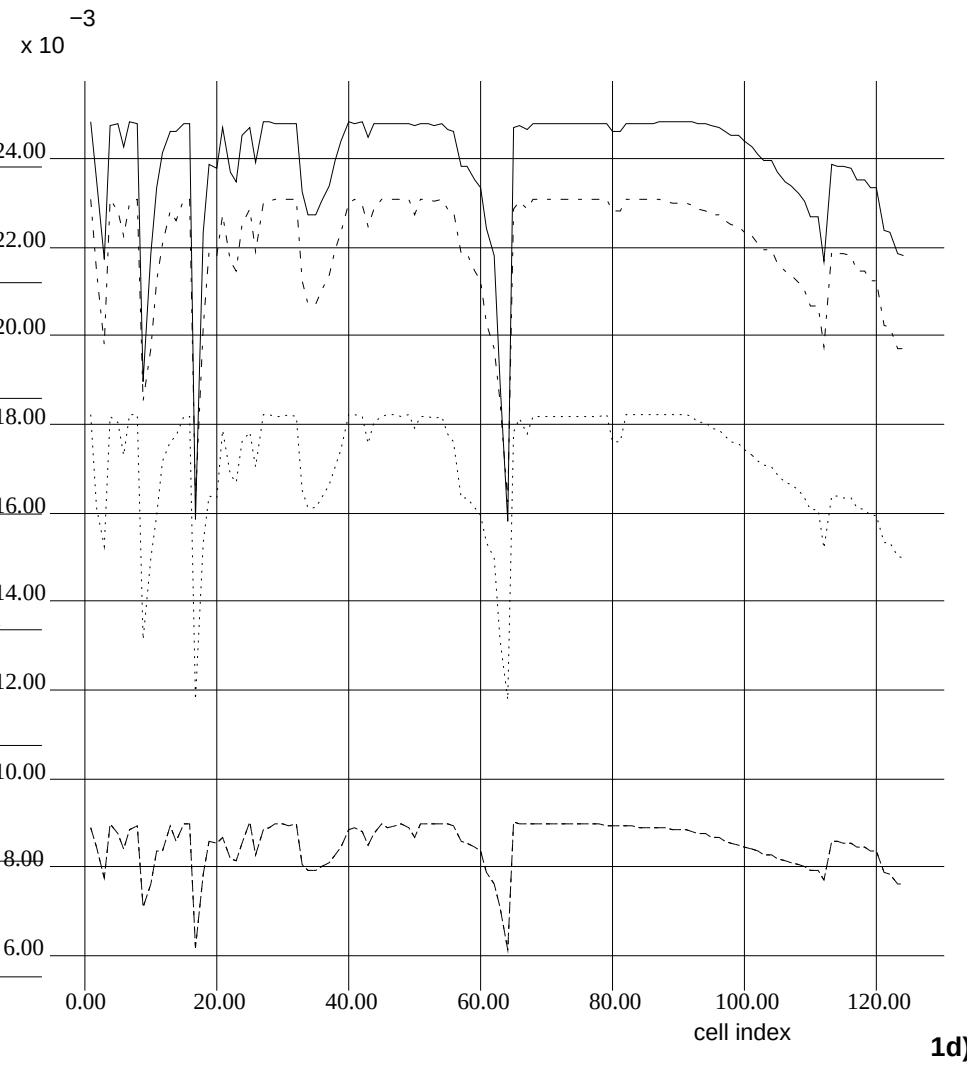

1d)

are two levels of differences by cells, one for the change of phases of metabolic oscillations, and the other for the fixed differentiation. Indeed this two-level differentiation gives a source for the hierarchical organization, since the phase difference within each group leads to further fixed differetiations later. It is also interesting to note that the phase difference is by "analogue" means, while the fixed differentiation of averages is discrete, where cells are separated "digitally".

**(4)Separation of inherent time scales**

When differentiated, two groups of cells show distinct orbits, which lie in different regimes in the $k$-dimensional phase space $\{x(j)\}$. In Fig.2, we have plotted overlaid orbits of two groups of cells. In each group, the oscillation phases are different by cells but the orbits fall in the same attractor, while the difference of orbits between the two groups is clearly discernible.

Another important feature here is the differentiation of the oscillation frequency. One group of cells oscillates faster than the other group. Typically cells with lower activities oscillate in time more slowly with smaller amplitudes. Thus inherent time scales differ by cells, and the division speeds of cells are also differentiated. Indeed, one group of cells divide faster than the other group of cells.

**(5) Transmission of Differentiation to Daughter Cells**

After the fixed differentiation, characteristic chemical compositions of each group are inherited by their daughter cells. Daughters of a cell of a given type keep the same character. Indeed, the cells with weaker activities in Fig.1, are successive daughters of an "ancestor" cell with such activity. In other words, when the system enters into this stage, a cell loses totipotency. By using the term in the cell biology [1], we call that the determination of a cell has occurred at this stage, since daughters of one type of cells preserve its type.

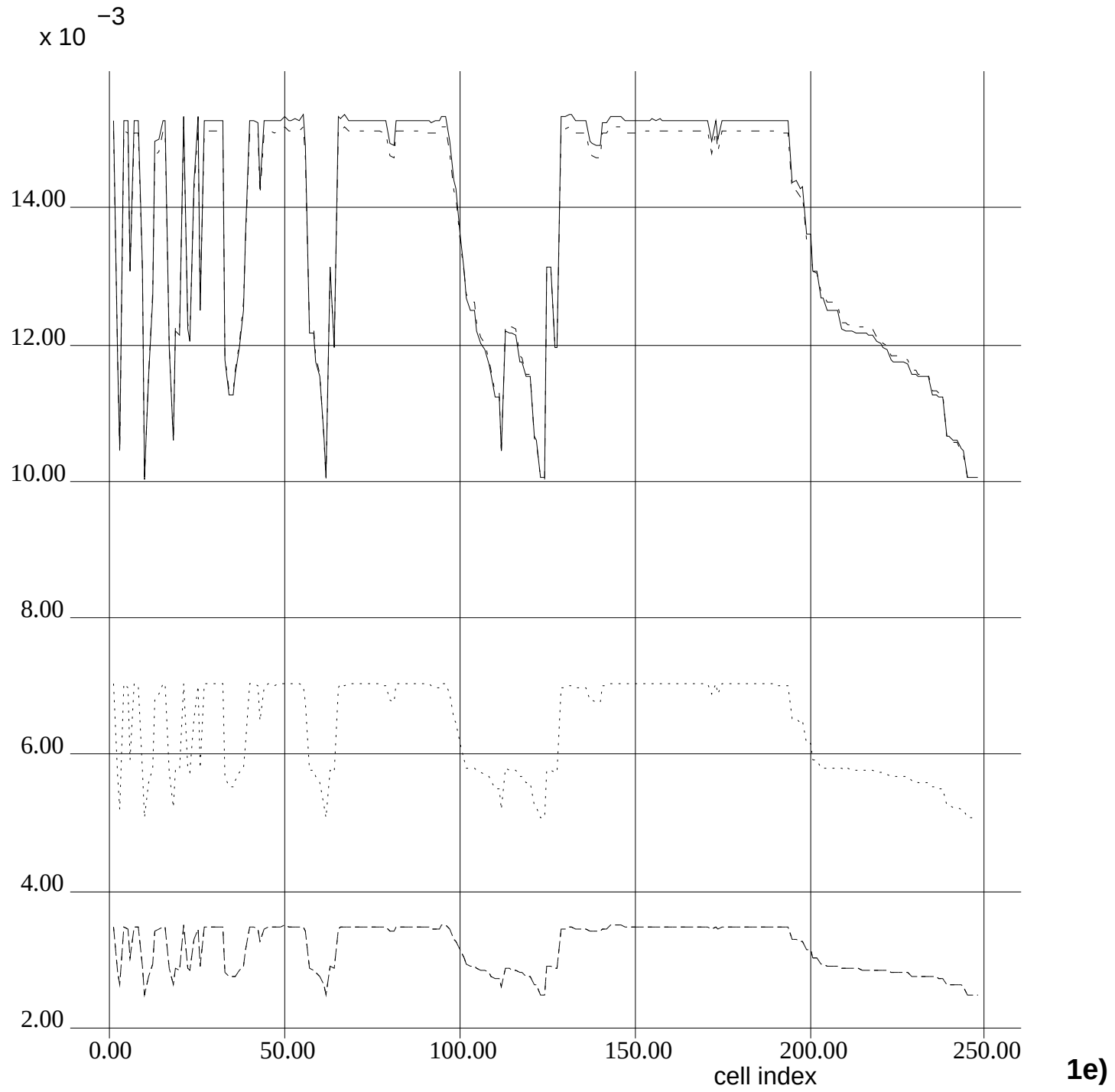

**Fig. 1.** Average chemical concentrations of $x^k(i)$. The cell index is defined in the order that the cell is born. Throughout the figures we use the parameters $e_1 = 1$, $s = 40$, $D = 0.02, f = 0.5$, $\delta = 0.2$, and $R = 100$, with the metabolic network of $k = 8$ chemicals, with 4 randomly chosen autocatalytic paths, although a variety of networks with the connection number from 2 to 4 lead to similar patterns of differentiation. The average is taken over the time steps while the cell number remains 16 (Fig.1a), 32 ( Fig.1b), 64 (Fig.1c), 124 ( Fig.1d), and 248 (Fig.1e). For reference, the snapshot values of two chemicals are also overlaid in Fig.1a).

It is important to note that the chemical characters are "inherited" just through the initial conditions of chemicals after the division, without any external implementation for a genetic transmission. The almost "digital" distinction of chemical characters, noted previously, is relevant to their preservation into daughter cells, since analogue differences may easily be disturbed by a possible noise at the division process.

**(6) Hierarchical Differentiation**

Further differentiation and determination of cells proceed successively in time. New types of cells are generated hierarchically. For example after two types of cells are differentiated ( let us call them type-A and type-B cells for the moment) at the first stage, the type-A cell is differentiated into A1 and A2. Here the difference between A1 and A2 ( for example that of chemicals or the frequency) is smaller than that between A and B. (see the two levels of "stronger" chemicals in Fig.1c)d)). Once this differentiation occurs, this character is fixed again, and after some time, such characters are determined by its daughter cells. With the cell divisions, this hierarchical determination of cells successively continues. For example daughters of type-B cells differentiate into B1 and B2 at a later stage. Examples of these successive determinations can be seen in Fig 1d)-e).

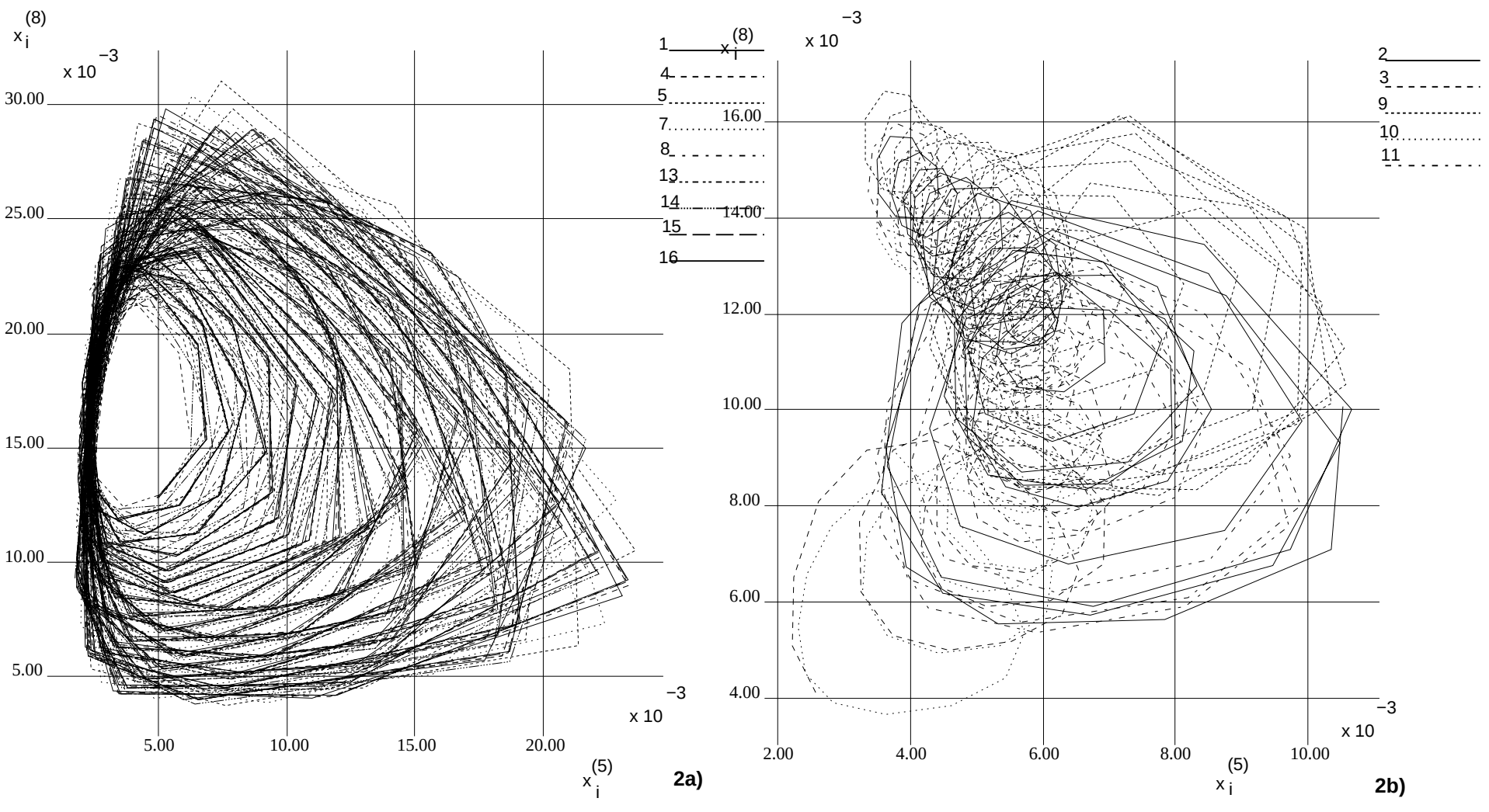

**Fig. 2.** Orbits of metabolic oscillations. Plotted are $(x_i^5(t), x_i^8(t))$ for time steps 1500 to 2000. ( The piecewisely-discontinuos character in the orbit is just an artifact in the plot, since the data are sampled by 0.5 second due to the limited memory of data space, and the orbit, of course, shows a smooth change). In Fig. 2a), overlaid are orbits for the cell index 1,4,5,7,8,13,14,15,16, which have same average chemical characters as seen in Fig.1. Note that the oscillation phase differs by cells. The orbits of the cell 2,3,9,10,11 are plotted in Fig.2b), where the difference between the "strong" type cell in Fig.2a) is clearly discernible ( also note the difference of scales).

Since the daughters of A (B) cells can be either A1 (B1) or A2 (B2) cells, the A cell is regarded as the stem cell over the later A1 and A2 cells. The daughters of A1 cells, on the other hand, remain to be A1-type cells, and may further differentiate into cells with smaller differences ( say A11 and A12 cells). In this case, the A1 cell can again be a stem cell over a narrower group of cells.

## 5 Further Remarks on Dynamic Differentiation

It is useful to make some remarks about the mechanism how the above scenario works and to give some possible predictions on the stability of differentiation processes.

**1) Initiation of the differentiation**

In our simulation, the differentiation starts after some divisions have occurred ( e.g., the number of cells becomes 16). Since the division leads to almost equal cells, a minor difference between the two cells must be enhanced to lead to a macroscopic differentiation. We have confirmed that a tiny difference of chemicals of very low concentrations is amplified to make a macroscopic difference of other chemicals with higher concentrations. It is interesting to note that such chemical with a low concentration is important, rather than those with high concentrations. This observation reminds us of a certain protein [1] that is known to have a signal transmission in order to trigger a switch to differentiation with only a small number of molecules.

**2) Stability of our scenario**

It should be noted that our scenario, although based on the chaotic instability, is rather robust against changes of initial conditions. Of course, which cell becomes one given type can depend on the initial conditions. The number distribution of each type of cells, on the other hand, is stable against a wide change of initial conditions. Still, if we start from a totally different type of initial conditions, such as those with many identical cells, the results can change as is discussed later.

**3) Memory as the inherited initial condition**

The differentiation in our scenario is originated in the interaction among cells, but the chemical characters of a cell are later memorized as the initial condition after the division. The differentiation with the interaction mechanism is reversible, while the latter mechanism leads to the determination. It is interesting to note that the determination is not implemented in the model in advance, but emerges spontaneously at some stage when the cell number $N$ exceeds some value.

In the natural course of the differentiation and in our simulations in §4, however, it is not possible to separate the memory in the inherited initial condition from the interaction with other cells. To see the tolerance of the memory as the inherited initial condition, one of the most effective methods is to pick up a determined cell and transplant it within a variety of surrounding cells, that are not seen in the "normal" course of the differentiation and development. Let us discuss some results of this "transplantation" experiments.

**4) Transplantation of cells**

In real biological experiments on the differentiation, some "artificial" initial conditions are adopted by transplantations of some types of cells. To check the validity of our scenario and to see the tolerance of the memory in the inherited initial condition, we have made several numerical experiments taking a variety of "artificial" initial conditions. As the initial conditions we choose determined cells at a later stage and mix them with undifferentiated cells at the earlier stage, to make the following observations.

*i) Starting from few determined cells of the same type in addition to undifferentiated cells*

The former group of cells keep its type, whose offsprings remain as the same type. Thus the determination is preserved, and the memory in the inherited initial conditions is robust against the change of cell interactions. The undifferentiated cells, on the other hand, start to differentiate to form many types of cells.

*ii) Starting only from few differentiated cells of the same type*

We have found either that the cells lose the non-trivial metabolic reaction, or that they start to differentiate again to generate different types of cells. Here the trivial reaction means that only one or two chemicals take non-zero values, without any ongoing reaction among the chemicals $\{x^j\}$ ( i.e., the only direct path of the *source* $x^0 \rightarrow x^m \rightarrow$ *the final product* is active). This type of cells with trivial reactions divides faster since the metabolic path way is direct. Taking also into account the fact that these cells destruct the chemical order sustained in the cell society, one may regarded them as tumor cells. The formation of these tumor cells depends on initial densities of determined cells, which may be compared with the experiments by Rubin [3].

## 6 Discussions and Biological Implications

To sum up we have proposed a novel scenario on the cell differentiation, based on the interacting cells with the metabolic oscillations and the clustering of coupled oscillator systems[5]. The model, without any external mechanism, leads to successive spontaneous differentiations, which are transmitted to daughter cells. It is interesting to note that a variety of experimental results can be understood from this point of view, such as the loss of totipotency, the origin of stem cells, the hierarchical differentiation, the separation of division speeds by the differentiation, the germ-line segregation, and the importance of chemicals of low concentrations for the trigger to the differentiation.

In the present paper we have not included a cell death process. When it is included, our model leads to a stationary distribution of differentiated cells at a later stage, when the cell number reaches its maximum[8]. Local interactions in space are not included either, to focus on the dynamic clustering process. The inclusion of a local interaction is rather straightforward, and indeed some simulations in a 2-dimensional space lead to the spatial organization of differentiated cells as well as the developmental process of some forms[8].

As a dynamics of many interacting agents, our results also provide novel interesting viewpoints. In particular we have succeeded in showing a mechanism of division of labors through the differentiation, as the segregation into active and inactive groups. It is interesting to extend the idea of the present paper to economics and sociology, and to discuss the origin of differentiation, diversity, and complexity there.

So far our results are rather universal as long as individual dynamics allows for some oscillations, and will be theoretically grounded by the studies of globally coupled dynamical systems, in particular, the spontaneous differentiation as the clustering [5]. Existence of a variety of chemicals in our problem leads to the "dual" clustering, both for the cell index and the chemical species. Construction of a minimal model with the dual differentiation will be an interesting problem as a dynamical systems theory, in future. Besides this viewpoint of coupled dynamical systems, it should be noted that our system is "open-ended" in the sense that the degrees of freedom increase with the cell division, where the notion of "open chaos" [6] will be useful to analyze the mechanism of the cell differentiations.

The authors would like to thank Chris Langton and Howard Gutowitz for useful discussions. This work is partially supported by a Grant-in-Aid for Scientific Research from the Ministry of Education, Science, and Culture of Japan. The authors would like to thank Chris Langton and the members of Santa Fe Institute for their hospitality during their stay, while the paper is completed.